\documentclass[
superscriptaddress,
preprintnumbers,
nofootinbib,
 amsmath,amssymb,
prd,
]{revtex4-2}
\usepackage{amsmath,amsthm}
\usepackage{amssymb,amsfonts}
\usepackage{bm}
\usepackage[colorlinks=true, linkcolor=blue, citecolor=blue, urlcolor=blue]{hyperref}
\usepackage{graphicx}
\usepackage{braket}
\usepackage{subfigure}
\usepackage{cases}
\usepackage{multirow}

\usepackage{algpseudocode}
\usepackage{fancyhdr}
\usepackage{setspace}

\allowdisplaybreaks

\newcommand{\kslash}{k\kern-1ex /}
\newcommand{\pslash}{p\kern-1ex /}
\newcommand{\qslash}{q\kern-1ex /}
\newcommand{\lslash}{l\kern-1ex /}
\newcommand{\sslash}{s\kern-1ex /}
\newcommand{\Dslash}{D\kern-1.2ex /}

\newcommand{\beqa}{\begin{eqnarray}}
\newcommand{\eeqa}{\end{eqnarray}}

\newcommand{\bd}{\begin{description}}
\newcommand{\ed}{\end{description}}

\newcommand{\eps}{\epsilon}

\newcommand{\ben}{\begin{eqnarray}}
\newcommand{\een}{\end{eqnarray}}
\newcommand{\nn}{\nonumber}
\def\lsim{\raise0.3ex\hbox{$<$\kern-0.75em\raise-1.1ex\hbox{$\sim$}}}
\def\gsim{\raise0.3ex\hbox{$>$\kern-0.75em\raise-1.1ex\hbox{$\sim$}}}
\def\simgt{\rlap{\lower 3.5 pt\hbox{$\mathchar \sim$}}\raise 2.0pt \hbox {$>$}}
\def\simlt{\rlap{\lower 3.5 pt\hbox{$\mathchar \sim$}}\raise 2.0pt \hbox {$<$}}

\usepackage{tikz,xcolor}
\definecolor{lime}{HTML}{A6CE39}
\DeclareRobustCommand{\orcidicon}{%
	\begin{tikzpicture}
	\draw[lime, fill=lime] (0,0) 
	circle [radius=0.16] 
	node[white] {{\fontfamily{qag}\selectfont \tiny ID}};	\draw[white, fill=white] (-0.0625,0.095) 
	circle [radius=0.007];	\end{tikzpicture}
	\hspace{-2mm}}
\foreach \x in {A,...,Z}{%
	\expandafter\xdef\csname orcid\x\endcsname{\noexpand\href{https://orcid.org/\csname orcidauthor\x\endcsname}{\noexpand\orcidicon}}
}

\begin{document}

\title{Tensor renormalization group study of cold and dense QCD in the strong coupling limit}
\author{Yuto Sugimoto\orcidA{}\,}
\email{sugimoto@nucl.phys.tohoku.ac.jp}
\affiliation{Department of Physics, Tohoku University, Sendai 980-8578, Japan}

\author{Shinichiro Akiyama\orcidB{}\,}
\email{akiyama@ccs.tsukuba.ac.jp}
\affiliation{Center for Computational Sciences, University of Tsukuba, Tsukuba, Ibaraki 305-8577, Japan}
\affiliation{Graduate School of Science, The University of Tokyo, Bunkyo-ku, Tokyo, 113-0033, Japan}

\author{Yoshinobu Kuramashi\orcidC{}\,}
\email{kuramasi@het.ph.tsukuba.ac.jp}
\affiliation{Center for Computational Sciences, University of Tsukuba, Tsukuba, Ibaraki 305-8577, Japan}
\begin{abstract}
We study the phase structure of the (3+1)-dimensional cold and dense QCD with the Kogut--Susskind quark in the strong coupling limit using the tensor renormalization group method. The chiral and nuclear transitions are investigated by calculating the chiral condensate and the quark number density as a function of the chemical potential. 
For a fixed temporal extent $N_\tau=8$, we determine the critical quark masses $m_c^{\chi}$ and $m_c^{n}$ for the chiral condensate and the quark number density, respectively, at which the first-order phase transition terminates with the vanishing discontinuity in thermodynamic quantities. 
We find that both quantities at the same quark mass exhibit a discontinuity at the same chemical potential, and the resulting critical quark masses are consistent with each other.
We also compare our results for the critical quark masses with those obtained from the Monte Carlo simulation in the dual formulation and from the mean-field analysis.
We further confirm the first-order phase transition at finite quark mass on a $1024^4$ lattice, which is essentially in the thermodynamic limit at zero temperature, as expected from the mean-field analysis. 
\end{abstract}
\date{\today}

\preprint{UTHEP-815, UTCCS-P-173}

\maketitle


\section{Introduction}
\label{sec:intro}

Theoretically, the cold and dense QCD is expected to exhibit a rich phase structure: As the density increases, a nuclear phase transition is followed by a chiral phase transition, and at the highest densities, color-superconducting and color-flavor locked phases may appear~\cite{Barrois:1977xd,Asakawa:1989bq,Alford:1997zt,Rapp:1997zu,Alford:1998mk}. 
Unfortunately, the cold and dense region of QCD is not accessible experimentally. 
Furthermore, lattice QCD simulations based on the Monte Carlo approach are hindered by the complex action problem at finite baryon chemical potential.
In this context, QCD in the strong coupling limit has been regarded as a useful starting point, since the theory can be reformulated in terms of dual variables~\cite{Karsch:1988zx,DelgadoMercado:2013ybm}, allowing Monte Carlo simulations to be performed in the dual formulation~\cite{Wolff:1984we,Fromm:2008ab,Miura:2009nu,deForcrand:2009dh,deForcrand:2014tha}.
The mean-field (MF) phase diagram has also been available~\cite{Kawamoto:1981hw,Bilic:1991nv,Nishida:2003fb,Nakano:2009bf}.
Recent studies, using not only the dual formulation~\cite{Kim:2023dnq} but also the complex Langevin method~\cite{Tsutsui:2021bog}, have been performed on relatively small lattice sizes up to $O(10^4)$ to tame the complex action problem.

Recently, we have applied the tensor renormalization group (TRG) method~\footnote{In this paper, ``TRG method'' refers not only to the original numerical algorithm proposed by Levin and Nave \cite{Levin:2006jai}, but also to its extensions~\cite{PhysRevB.86.045139,Shimizu:2014uva,PhysRevLett.115.180405,Sakai:2017jwp,PhysRevLett.118.110504,Hauru:2017tne,Adachi:2019paf,Kadoh:2019kqk,Akiyama:2020soe,PhysRevB.105.L060402,Akiyama:2022pse}.} to the (3+1)-dimensional ((3+1)$d$) dense two-color QCD (QC$_2$D) at zero temperature on a $1024^4$ lattice in the strong coupling limit, whose phase structure was investigated by measuring the chiral and diquark condensates and the quark number density as a function of the chemical potential~\cite{Sugimoto:2025vui}. The critical exponents for the diquark condensation were also determined. Since the application of the TRG method to the non-Abelian gauge theory with dynamical fermions is restricted to the two-dimensional case due to the rapid growth of the degree of freedom in terms of dimensionality~\cite{Bloch:2022vqz,Asaduzzaman:2023pyz,Pai:2024tip,Pai:2025eia}, our previous work has served as a preparatory study before exploring the (3+1)$d$ QCD.     

In this paper, building on our previous study, we now extend the TRG method to the (3+1)$d$ QCD with the staggered quark at finite density in the strong coupling limit, following the computational strategy employed in our previous study of the (3+1)$d$ QC$_2$D. 
For the TRG method, this constitutes a technical extension from two to three colors, regardless of the negative sign problem and complex action problem.
In fact, the TRG method has been successfully applied to various models with such problems to investigate their phase structures~\cite{Shimizu:2014uva,Shimizu:2014fsa,Kawauchi:2016xng,Kawauchi:2017dnj,Yang:2015rra,Shimizu:2017onf,Takeda:2014vwa,Kadoh:2018hqq,Kadoh:2019ube,Takeda:2019idb,Butt:2019uul,Kuramashi:2019cgs,Akiyama:2020ntf,Akiyama:2020soe,Nakayama:2021iyp,Bloch:2021mjw,Bloch:2022vqz,Akiyama:2023hvt,Akiyama:2024qer,Luo:2022eje,Hite:2024ulb,Kanno:2024elz,Luo:2024lbh,Luo:2025qtv,Aizawa:2025lxi}.  
We first investigate the chiral and nuclear transitions at $N_\tau=8$, where $N_{\tau}$ denotes the temporal lattice size, using the chiral condensate and the quark number density. 
Their critical endpoints $m_c^{\chi}$ and $m_c^{n}$, at which the first-order phase transition terminates, are determined by locating the points where the discontinuities in the chiral condensate and the quark number density vanish, respectively.
We also compare the resulting critical quark masses with those obtained from the Monte Carlo simulation in the dual formulation~\cite{Kim:2023dnq} and from the MF analysis~\cite{Nishida:2003fb}.
We then study the first-order phase transition at finite quark mass in the vanishing temperature limit, as predicted by the MF approach~\cite{Nishida:2003fb}, by employing a large lattice of size $1024^4$.

This paper is organized as follows. In Sec.~\ref{sec:method}, we define the action of the strong coupling QCD with finite chemical potential and give the Grassmann tensor network representation. 
In Sec.~\ref{sec:results}, we determine the critical endpoints $m_c^{\chi,n}$ at $N_\tau=8$ by measuring the chiral condensate and the quark number density. 
We also confirm the first-order phase transition at finite quark mass in the thermodynamic limit at zero temperature.
Section~\ref{sec:summary} is devoted to summary and outlook.

\section{Formulation and numerical algorithm}
\label{sec:method}
\subsection{Strong coupling QCD}
Although the formulation of the $(3+1)d$ finite density QCD at the strong coupling limit is straightforwardly available by replacing $N_c=2$ by $N_c=3$ in the formulation for the QC$_2$D case given in Ref.~\cite{Sugimoto:2025vui}, we again present it to make this paper self-contained. 
  
The finite density QCD is considered on a (3+1)$d$ lattice $\Lambda_{3+1}=\{(n_1,n_2,n_3,n_4)\ \vert n_{1,2,3}=1,\dots,N_s\, , n_4=1,\dots,N_\tau\}$ whose volume is $V=N_s^3\times N_\tau$ and temperature is defined as $T=1/N_\tau$. The lattice spacing $a$ is set to $a=1$ unless necessary. 
We employ the Kogut--Susskind quark action with the finite chemical potential $\mu$:
\begin{align}
  S_F=\sum_{n\in\Lambda_{3+1}}\sum_{\nu=1}^{4} \frac{\eta_\nu(n)}{2}\left[e^{\mu\delta_{\nu,4}}{\bar \chi(n)}U_\nu(n)\chi(n+{\hat \nu})- e^{-\mu\delta_{\nu,4}}{\bar \chi(n+{\hat \nu})}U_\nu^\dagger(n)\chi(n)\right]
  +m\sum_{n}{\bar \chi(n)}\chi(n),
  \label{eq:action_f}
\end{align}
with $m$ the quark mass. 
The $SU$($N_{c}$)-valued $U_\nu(n)$ is introduced on the link between the sites $n$ and $n+\hat{\nu}$, with $\hat{\nu}$ the unit vector in the $\nu$ direction. 
Quark fields are denoted by $N_c$-component Grassmann variables $\chi=(\chi^1,\chi^2,\dots,\chi^{N_c})$ and $\bar{\chi}=(\bar{\chi}^1,\bar{\chi}^2,\dots,\bar{\chi}^{N_c})$, and $\eta_\nu(n)$ is the staggered sign function defined by $\eta_\nu(n)=(-1)^{n_1+\cdots +n_{\nu-1}}$ with $\eta_1(n)=1$.
Note that Eq.~(\ref{eq:action_f}) is invariant under the following continuous chiral transformation at $m=0$:
\begin{align}
    \chi(n)\rightarrow e^{i\alpha\epsilon(n)}\chi(n),
    ~~~
    \bar{\chi}(n)\rightarrow \bar{\chi}(n)e^{i\alpha\epsilon(n)},
\end{align}
with $\alpha \in \mathbb{R}$ and $\epsilon(n)=(-1)^{n_1+n_2+n_3+n_4}$.
 
The partition function in the strong coupling limit is defined as
\begin{equation}\label{eq:partition_function}
  Z=\int \left(\prod_{n\in\Lambda_{3+1}}d\chi(n)d\bar{\chi}(n)\right)
  \left(\prod_{n}\prod_{\nu=1}^{4}dU_\nu(n)\right)
  e^{-S_F}.
\end{equation}
The gauge action is absent from the partition function since $1/g^{2}=0$, and the link variables remain only in the quark sector.

\subsection{Grassmann tensor network representation}
We now explain how to derive a tensor network representation, following the Grassmann tensor network formulation provided in Ref.~\cite{Akiyama:2020sfo}. 
The partition function in Eq.~\eqref{eq:partition_function} is composed of local Boltzmann weights assigned to every link, and the Grassmann variables $\chi$ and $\bar{\chi}$ are located at each lattice site. 
We insert $N_c$-component auxiliary Grassmann variables $\zeta=(\zeta^1,\dots,\zeta^{N_c})$, $\xi=(\xi^1,\dots,\xi^{N_c})$ and their conjugate $\bar{\zeta}$, $\bar{\xi}$ into the hopping term of original quark fields as
\begin{align}
&\exp\left[-\frac{\eta_\nu(n)}{2}e^{\mu\delta_{\nu,4}}\bar{\chi}(n)U_\nu(n){\chi}(n+\hat{\nu})\right] \nonumber\\
&= \int \left[\prod_{i=1}^{N_c}d\bar{\zeta}_\nu^i(n)d\zeta_\nu^i(n)e^{-\bar{\zeta}_\nu^i(n)\zeta_\nu^i(n)}\right]\exp\bigg[-\frac{\eta_\nu(n)}{2}e^{\mu\delta_{\nu,4}}\bar{\chi}(n)\zeta_\nu(n)+\bar{\zeta}_\nu(n)U_\nu(n)\chi(n+\hat{\nu})\bigg],
\end{align}
\begin{align}
&\exp\left[\frac{\eta_\nu(n)}{2}e^{-\mu\delta_{\nu,4}}\bar{\chi}(n+\hat{\nu})U^\dagger_\nu(n)\chi(n)\right] \nonumber\\
&= \int \left[\prod_{i=1}^{N_c}d\bar{\xi}_\nu^i(n)d\xi_\nu^i(n)e^{-\bar{\xi}_\nu^i(n)\xi_\nu^i(n)}\right]\exp\left[\bar{\chi}(n+\hat{\nu})U^\dagger_\nu(n)\bar{\xi}_\nu(n)-\frac{\eta_\nu(n)}{2}e^{-\mu\delta_{\nu,4}}\xi_\nu(n)\chi(n)\right].
\end{align}
We then define the local Grassmann tensor $\mathcal{T}$ at the lattice site $n$ by integrating out $\chi(n),\,\bar{\chi}(n)$ and $U_\nu(n)$ as follows,
\begin{align}\label{eq:local_tensor}
\mathcal{T}&=\int \left(\prod_{\nu=1}^{4} dU_\nu\right)\left(\prod_{i=1}^{N_c}d\chi^id\bar{\chi}^ie^{-m\bar{\chi}^i\chi^i}\right)\nn\\
&\qquad\times \exp\bigg[-\frac{\eta_\nu(n)}{2}e^{\mu\delta_{\nu,4}}\bar{\chi}\zeta_\nu(n)+\bar{\zeta}_\nu(n-\hat{\nu})U_\nu(n-\hat{\nu})\chi\bigg]\nn\\
&\qquad\times\exp\bigg[\bar{\chi}U^\dagger_\nu(n-\hat{\nu})\bar{\xi}_\nu(n-\hat{\nu})-\frac{\eta_\nu(n)}{2}e^{-\mu\delta_{\nu,4}}\xi_\nu(n)\chi\bigg].
\end{align}
This Grassmann tensor forms a Grassmann tensor network representing the partition function $Z$ as
\begin{align}
\label{eq:Gtn}
    Z=\mathrm{gTr}\left[\prod_{n}\mathcal{T}_{\Psi_1 (n)\Psi_2 (n)\Psi_3 (n)\Psi_4 (n)\,
    \bar{\Psi}_4 (n-\hat{4})\bar{\Psi}_3 (n-\hat{3})\bar{\Psi}_2 (n-\hat{2})\bar{\Psi}_1 (n-\hat{1})}\right],
\end{align}
where we defined composite auxiliary Grassmann variables $\Psi_\nu=(\zeta^1_\nu,\zeta^2_\nu,\zeta^3_\nu,\xi^1_\nu,\xi^2_\nu,\xi^3_\nu)$ and 
$\bar{\Psi}_\nu=(\bar{\zeta}^1_\nu,\bar{\zeta}^2_\nu,\bar{\zeta}^3_\nu,\bar{\xi}^1_\nu,\bar{\xi}^2_\nu,\bar{\xi}^3_\nu)$ and $\mathrm{gTr}$ denotes the Grassmann trace, which is defined by $\prod_{n,\nu}\int\int d\bar{\Psi}_\nu(n)d\Psi_\nu(n)e^{-\Psi_\nu(n)\bar{\Psi}_\nu(n)}$. We impose the periodic boundary condition for $\hat{1},\hat{2},\hat{3}$ directions, and antiperiodic boundary condition for the $\hat{4}$ direction.
The explicit formula for coefficient tensor of $\mathcal{T}$ is described in Appendix~\ref{app:Gtn}.
The local Grassmann tensor in Eq.~\eqref{eq:local_tensor} consists of eight types of tensors arising from its staggered phase. 
The network is uniform along the $\hat{4}$ direction, while it exhibits periodic structures in the $\hat{1}$, $\hat{2}$, and $\hat{3}$ directions~\cite{Akiyama:2020soe}. 

To evaluate Eq.~\eqref{eq:Gtn}, we employ the anisotropic TRG (ATRG) algorithm~\cite{Adachi:2019paf}, accelerated by the bond-swapping technique~\cite{Oba:2019csk}, whose efficacy in approximating the contraction of Grassmann tensor networks has already been established in Ref.~\cite{Akiyama:2020soe}.
In this work, the Grassmann ATRG algorithm is further accelerated by the multi-GPU parallelization strategy detailed in Ref.~\cite{Sugimoto:2025xva}.
For finite-temperature calculations, we first perform the four-dimensional Grassmann ATRG coarse-graining with the bond dimension $D$ along the $\hat{4}$ direction, repeating the procedure $\log_2{N_\tau}$ times for each type of initial tensor.
We then trace out the $\hat{4}$ direction with antiperiodic boundary conditions, and obtain a three-dimensional cubic Grassmann tensor network, which is subsequently evaluated by applying the three-dimensional Grassmann ATRG with the bond dimension $D_{3d}$
\footnote{
The coarse-graining along the $\hat{4}$ direction corresponds to the imaginary-time evolution.
For anisotropic systems, it is known from previous TRG studies that performing the imaginary-time evolution first leads to a more efficient approximate contraction~\cite{PhysRevB.86.045139,Akiyama:2021xxr,Akiyama:2021glo}.
Furthermore, a recent study using the thermal tensor network renormalization method~\cite{Ueda:2025mhu} has also reported the advantage of performing the imaginary-time evolution first to reduce the dimensionality of the tensor network.
}.
For zero-temperature calculations, we first perform coarse-graining along the spatial directions to reduce the staggered unit cell to a single tensor.
We then iteratively apply the coarse-graining procedure in each direction in sequence, following the approach used in Ref.~\cite{Akiyama:2020soe} for the staggered Grassmann tensor network contraction.

\section{Numerical results}
\label{sec:results}

\subsection{Critical endpoint at $N_\tau=8$}
\label{subsec:nt=8}

We first check the convergence behavior of the thermodynamic potential,
\begin{align}
    f(m,\mu,D,D_{3d})=\frac{1}{V}\ln Z(m,\mu,D,D_{3d}),
\end{align}
by defining relative errors as 
\begin{align}
    \delta_{3d} f(m,\mu,D_{3d})=\left\vert \frac{\ln Z(m,\mu,D=55,D_{3d})-\ln Z(m,\mu,D=55,D_{3d}=125)}{\ln Z(m,\mu,D=55,D_{3d}=125)}\right\vert,
\end{align}
\begin{align}
    \delta f(m,\mu,D)=\left\vert \frac{\ln Z(m,\mu,D,D_{3d}=120)-\ln Z(m,\mu,D=60,D_{3d}=120)}{\ln Z(m,\mu,D=60,D_{3d}=120)}\right\vert.
\end{align}

\begin{figure}[htbp]
	\centering
	\includegraphics[width=0.7\hsize]{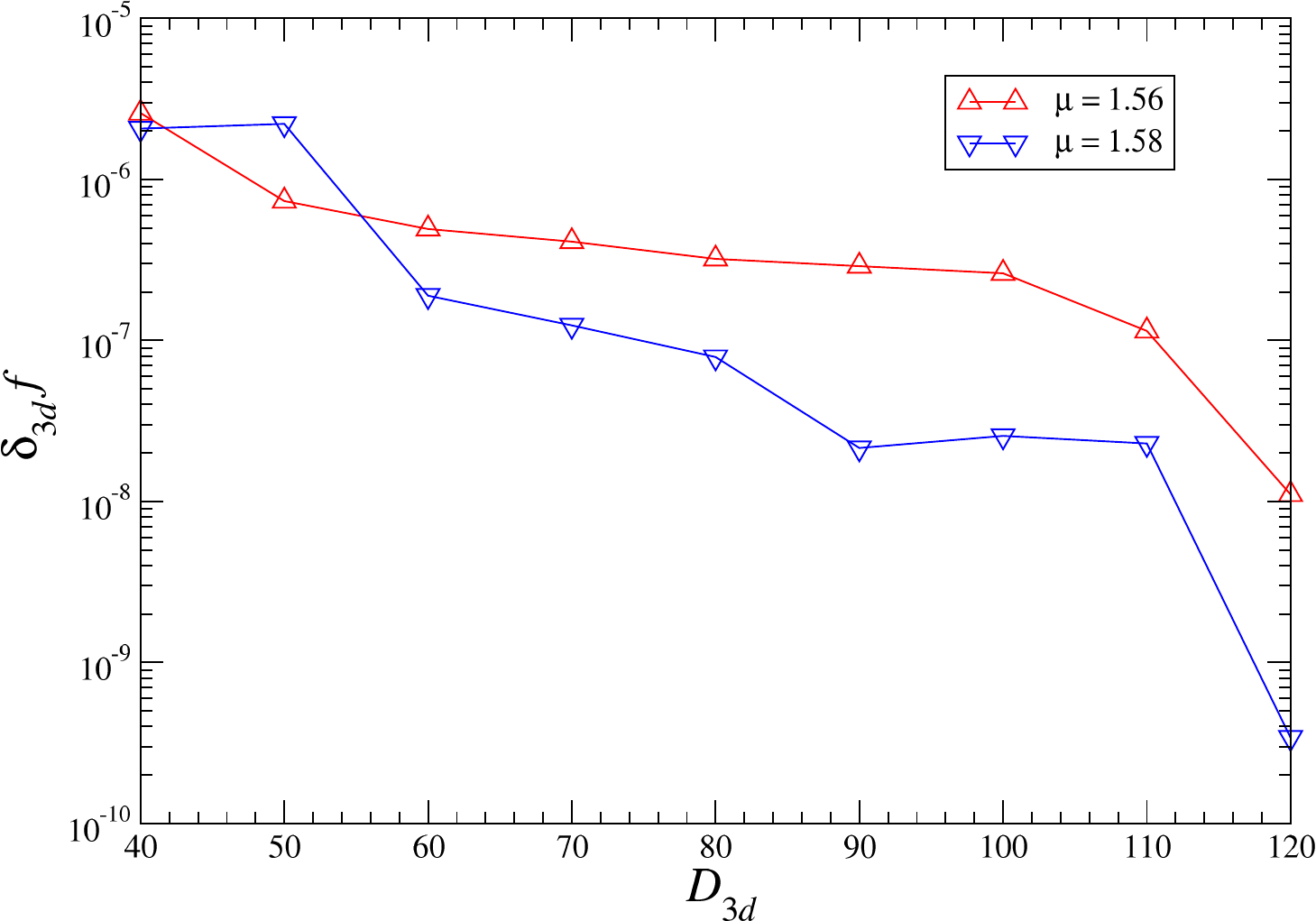}
	\caption{
        Relative error of thermodynamic potential in terms of $D_{3d}$ at $m=2.00$, for $\mu=1.56$ and 1.58, on a $32^3\times 8$ lattice.  
    }
  	\label{fig:delta_3d_f}
\end{figure}

\begin{figure}[htbp]
	\centering
	\includegraphics[width=0.7\hsize]{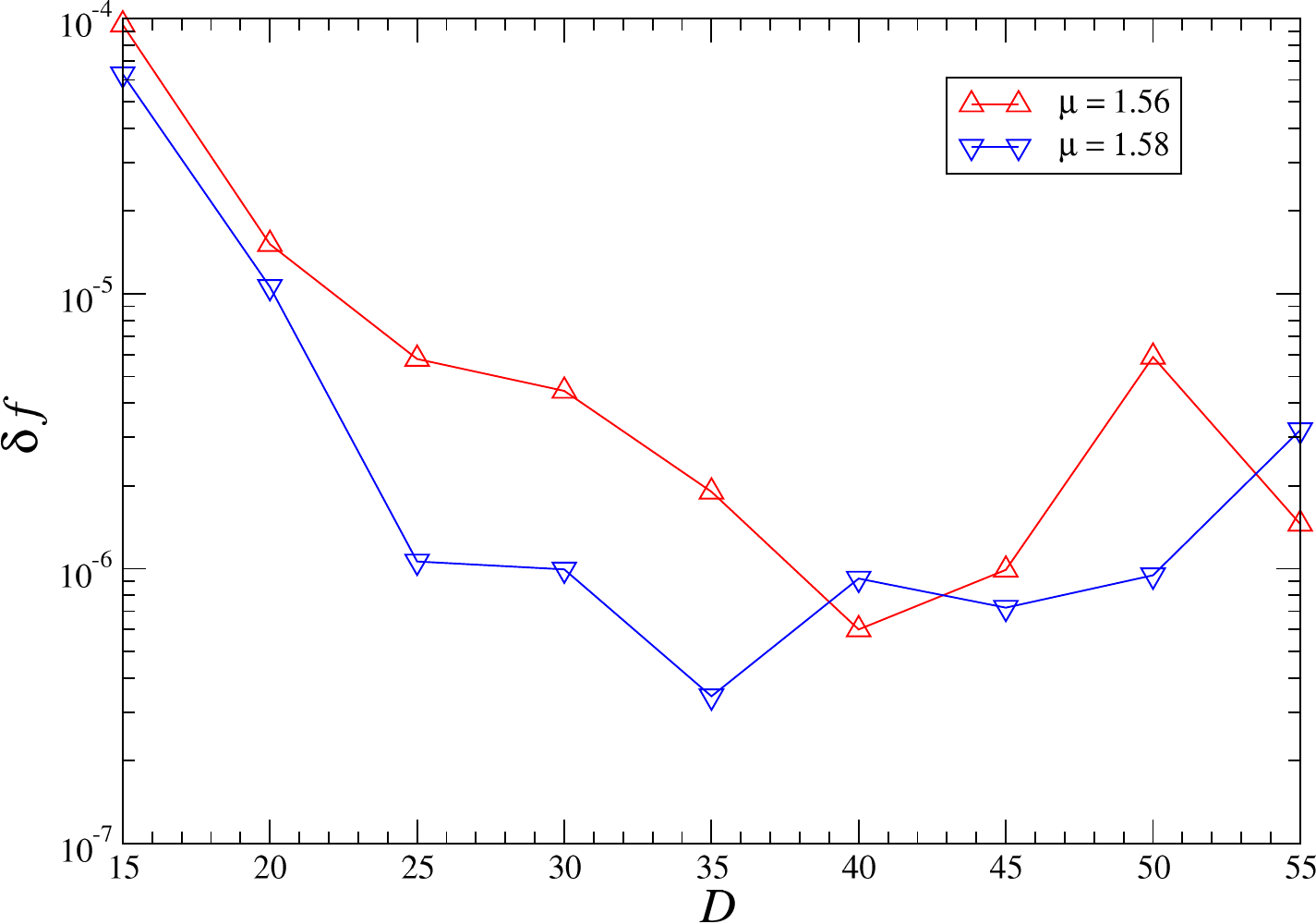}
	\caption{
        Relative error of thermodynamic potential in terms of $D$ at $m=2.00$, for $\mu=1.56$ and 1.58, on a $32^3\times 8$ lattice.  
    }
  	\label{fig:delta_f}
\end{figure}

Figure~\ref{fig:delta_3d_f} shows the $D_{3d}$ dependence of $\delta_{3d} f$ choosing $\mu=1.56$ and 1.58 at $m=2.00$, which is close to the critical endpoint discussed later. Both cases show good convergence behaviors: $\delta_{3d} f$ is less than $10^{-5}$ at $D_{3d}=120$ and $\delta f$ is $O(10^{-6})$ at $D=55$ as observed in Fig.~\ref{fig:delta_f}.
This is the reason why we used $D=55$ and $D_{3d}=120$ in defining $\delta_{3d}f$ and $\delta f$, respectively, where the relative errors can be evaluated at a sufficiently converged level for the following calculations.
Hereafter, we present the results obtained with $(D,D_{3d})=(55,120)$ and omit the explicit dependence on $D$ and $D_{3d}$ in the thermodynamic quantities.


\begin{figure}[htbp]
	\centering
	\includegraphics[width=0.9\linewidth]{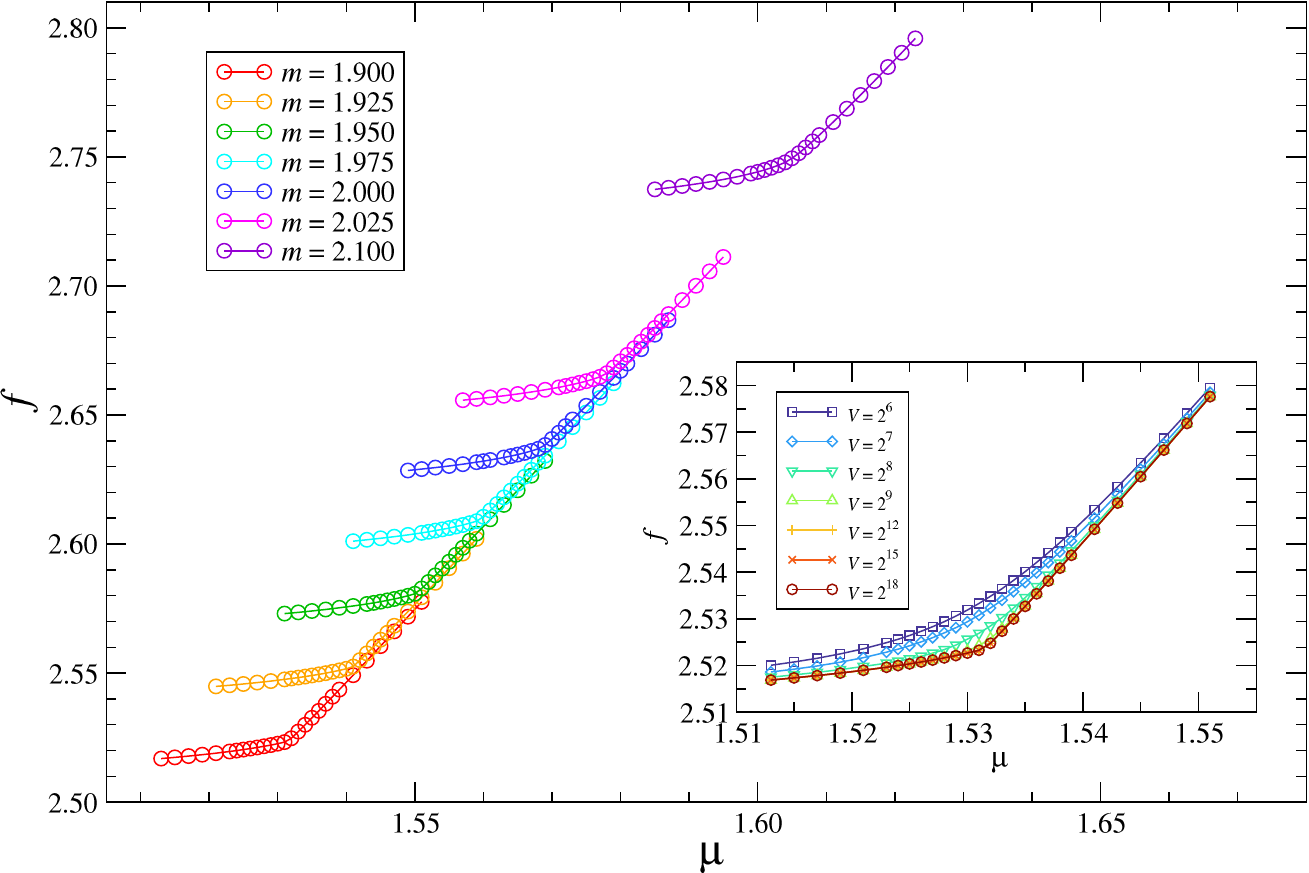}
	\caption{$\mu$ dependence of $f(m,\mu)$ on a $32^3\times 8$ lattice for $m\in[1.90,2.025]$. The inset figure shows the $\mu$ dependence of $f(m=1.9,\mu)$ for various volumes.}
  	\label{fig:f_nt8}
\end{figure}

\begin{figure}[htbp]
	\centering
	\includegraphics[width=0.85\hsize]{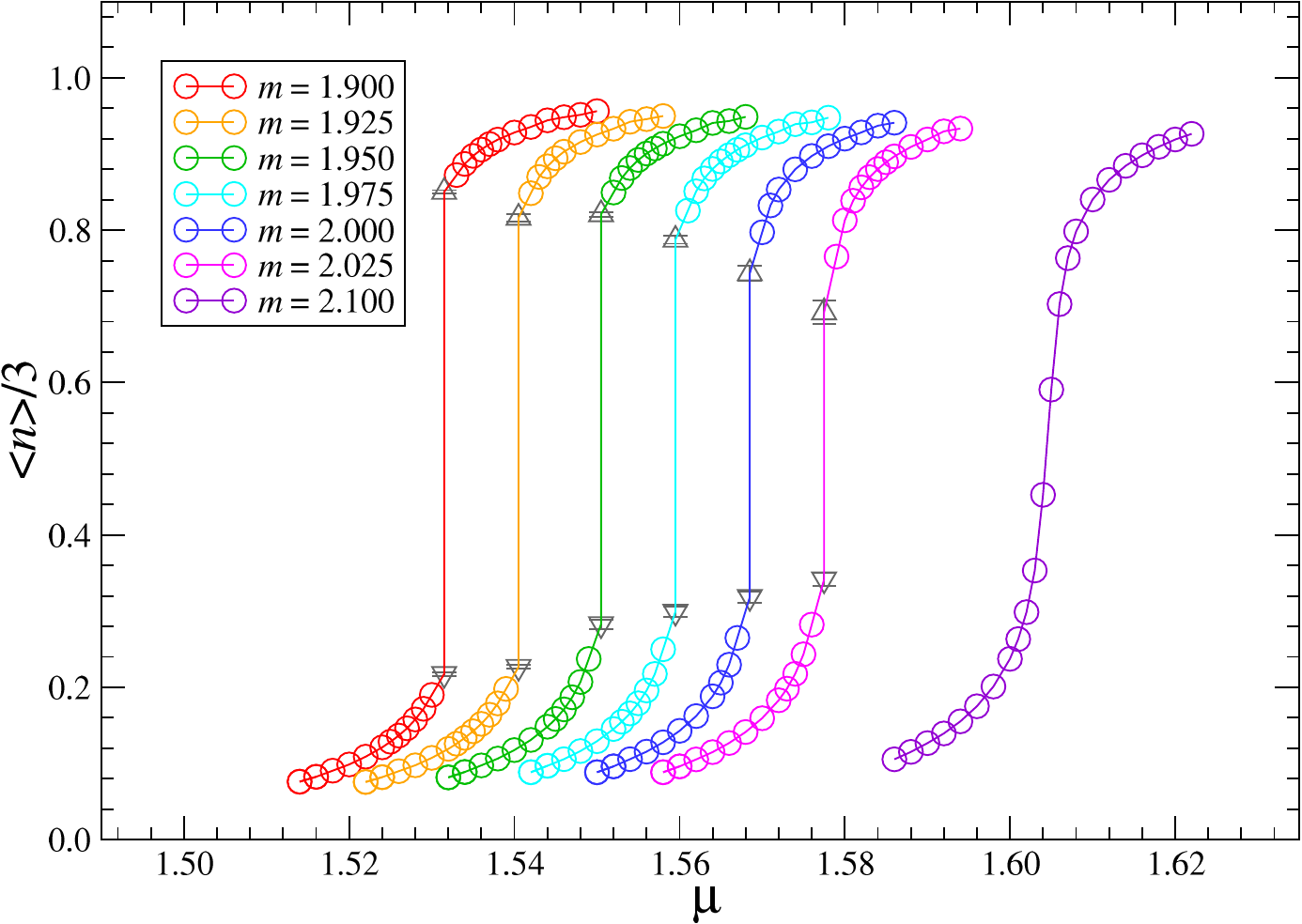}
	\caption{
        $\mu$ dependence of $\langle n \rangle(m,\mu)$ at $N_\tau=8$ with $m\in[1.90,2.025]$. 
        Triangles denote $\langle n \rangle (m,\mu\rightarrow (\mu_++\mu_-)/2\pm 0)$ (see main text for details). 
        The data at $m=2.10$ is also plotted for comparison.
    }
  	\label{fig:gap_nt8_n}
\end{figure}

We now discuss the nuclear transition. 
We evaluate the quark number density by the numerical derivative of the thermodynamic potential at $N_\tau=8$:
\begin{align}
\langle n \rangle(m,\mu) = \frac{1}{V}\frac{\partial \ln Z(m,\mu)}{\partial \mu}\simeq \frac{1}{V}\frac{\ln Z(m,\mu+\Delta \mu)-\ln Z(m,\mu-\Delta \mu)}{2\Delta \mu},
  \label{eq:<n>}
\end{align}
with $\Delta \mu=0.001$. 
This method, however, can cause a problem near the first-order transition point: There exists a possibility that the difference in the thermodynamic potentials across the transition point is measured. 
As a result, with a finite $\Delta \mu$ in Eq.~\eqref{eq:<n>}, the discontinuity of thermodynamic quantities associated with a first-order phase transition may be masked.
Figure~\ref{fig:f_nt8} shows the $\mu$ dependence of the thermodynamic potential with several choices of $m$. We observe a kink, which is supposed to give a discontinuity in the quark number density, at a certain value of $\mu$ for each $m\in[1.90,2.025]$.  
We then evaluate the magnitude of discontinuity $\Delta \langle n\rangle(m)$ as $\vert f(m,\mu\rightarrow (\mu_++\mu_-)/2+0)-f(m,\mu\rightarrow (\mu_++\mu_-)/2-0)\vert$, where the kink position is sandwiched between $\mu_+$ and $\mu_-$ with $\vert \mu_+-\mu_-\vert =0.001$. 
Here, the values of the thermodynamic potentials $f(m,\mu\rightarrow (\mu_++\mu_-)/2\pm 0)$ are obtained by fitting four data points closest to the kink position with a quadratic function in each phase. 
The resulting $\Delta \langle n \rangle(m)$ are summarized in Table~\ref{tab:gap_nt8}, where $\mu_c^{n}=(\mu_++\mu_-)/2$ can be regarded as a transition point, and the quoted error corresponds to $\vert\mu_+-\mu_-\vert$. 
We show the $\mu$ dependence of $\langle n\rangle(m,\mu)$ of Eq.~(\ref{eq:<n>}) in Fig.~\ref{fig:gap_nt8_n}, where the data on the kink position is omitted. 
Since the magnitude of discontinuity $\Delta \langle n\rangle(m)$ gradually shrinks as the quark mass increases, this behavior indicates the possible existence of a critical endpoint $m_c^{n}(N_\tau=8)$. 

\begin{figure}[htbp]
	\centering
	\includegraphics[width=0.85\hsize]{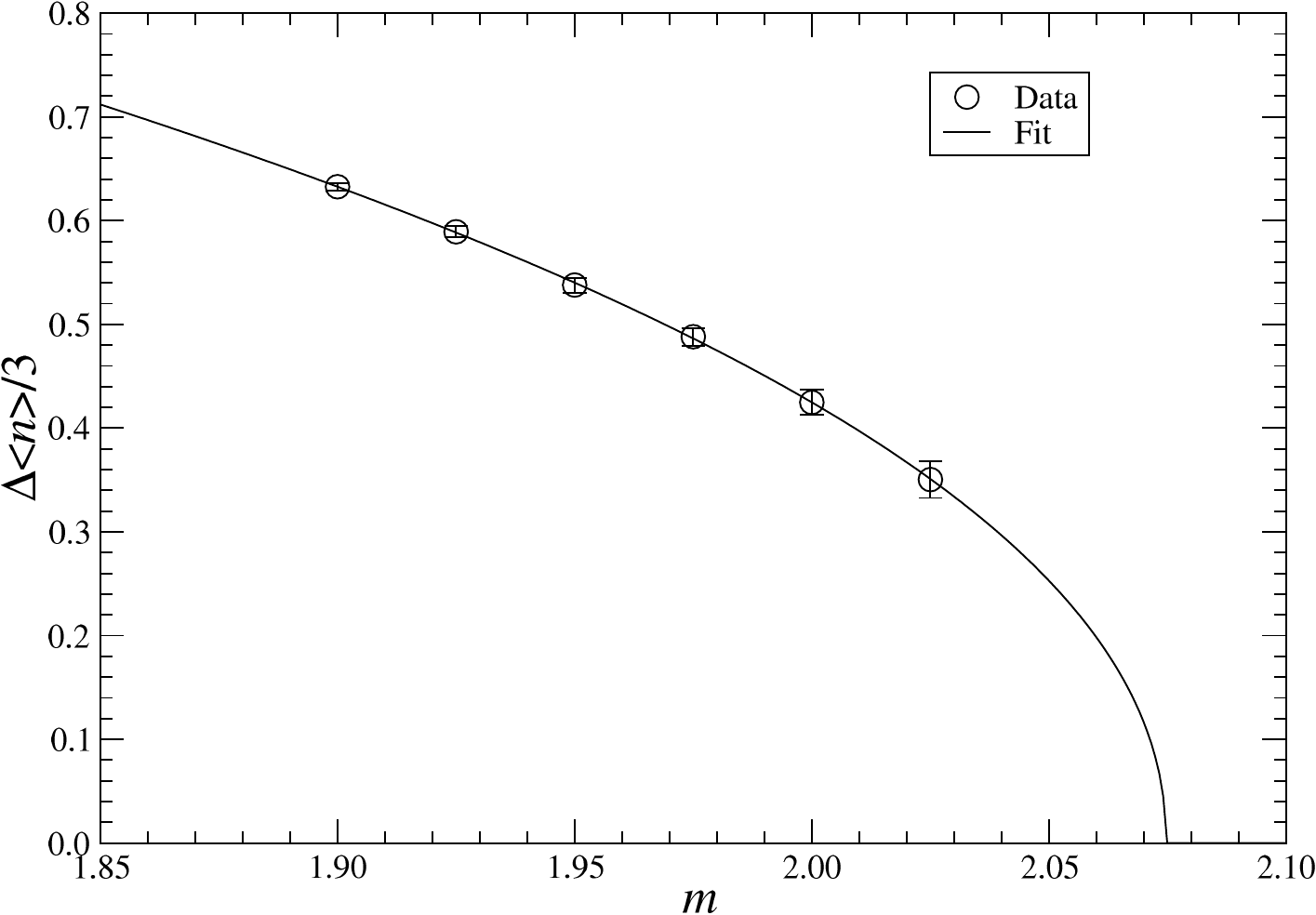}
	\caption{Fit of $\Delta \langle n \rangle(m)$ as a function of $m$. Solid curve denotes the fit result.}
  	\label{fig:fit_n}
\end{figure}

Figure~\ref{fig:fit_n} shows the resulting gap $\Delta \langle n\rangle(m)$ as a function of $m$. 
The solid curve represents the fit to $\Delta \langle n \rangle(m)$ assuming the functional form,
\begin{align}
    \Delta \langle n \rangle(m)=A_{n}\left(m_c^{n}(N_\tau=8)-m\right)^{p_{n}},
\end{align}
with $A_{n}$, $m_c^{n}(N_\tau=8)$, and $p_{n}$ are treated as free parameters. 
We obtain 
\begin{align*}
    A_{n}=1.43(16),~~~m_c^{n}(N_\tau=8)=2.075(23),~~~p_{n}=0.47(10).
\end{align*}
Our result for $m_c^n(N_\tau=8)$ is larger than the value $m_c^{n}(N_\tau=8)=1.7(1)$ estimated from the dual formulation~\cite{Kim:2023dnq}, while it is smaller than the mean-field prediction $m_c^{n}(N_\tau=8)\approx 2.4$~\cite{Nishida:2003fb,Kim:2023dnq}. 
For comparison, we also plot the data at $m=2.10$, which is expected to lie in the crossover region, in Figs.~\ref{fig:f_nt8} and \ref{fig:gap_nt8_n}. 
In this case, it is difficult to identify a jump in the $\mu$ dependence of $\langle n\rangle$.

\begin{table}[htb]
	\caption{
        Values of $m$, $\mu_c^n$, $\Delta \langle n\rangle(m)$, $\mu_c^\chi$ and $\Delta \langle {\bar \chi}\chi\rangle(m)$ at the first-order phase transition. 
        All the results are obtained with $(D,D_{3d})=(55,120)$.
    }
	\label{tab:gap_nt8}
	\begin{center}
	  	\begin{tabular}{|c|c|l|c|l|}\hline

			$m$ & $\mu_c^n$ & $\Delta \langle n\rangle$ & $\mu_c^\chi$ & $\Delta \langle {\bar \chi}\chi\rangle$ \\ \hline
			1.900 & 1.5315(5) & 0.6326(34)  & 1.532(1) & 0.6925(13)\\
            1.925 & 1.5405(5) & 0.589(5)  & 1.541(1) & 0.6388(15)\\
			1.950 & 1.5505(5) & 0.538(7)  & 1.550(1) & 0.5703(21)\\
            1.975 & 1.5595(5) & 0.488(8)  & 1.559(1) & 0.473(5)\\
			2.000 & 1.5685(5) & 0.425(12)  & 1.569(1) & 0.390(6)\\
            2.025 & 1.5775(5) & 0.350(18)  & 1.578(1) & 0.308(6)\\\hline

	\end{tabular}
	\end{center}
\end{table}

\begin{figure}[htbp]
	\centering
	\includegraphics[width=0.85\hsize]{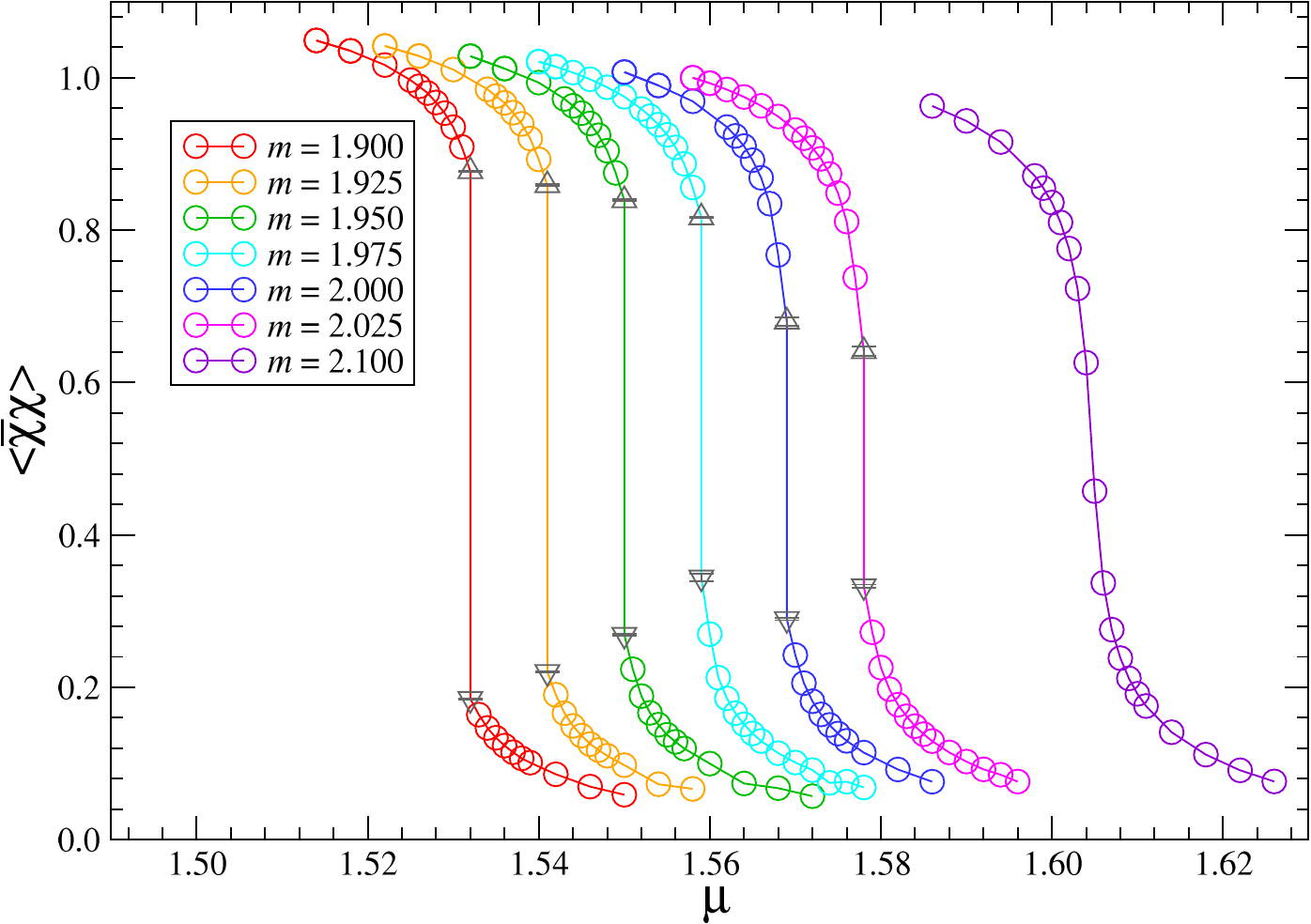}
	\caption{
        $\mu$ dependence of $\langle {\bar \chi} \chi \rangle(m,\mu)$ at $N_\tau=8$ with $m\in[1.90,2.025]$. 
        Triangles denote $\langle {\bar \chi} \chi \rangle (m,\mu\rightarrow (\mu_++\mu_-)/2\pm 0)$ (see main text for details). 
        The data at $m=2.10$ is also plotted for comparison.
        }
  	\label{fig:gap_nt8_chi}
\end{figure}

Let us turn to the chiral transition. We calculate the chiral condensate $\langle {\bar \chi} \chi\rangle$ at $N_\tau=8$ by the numerical derivative of the thermodynamic potential:
\ben
    \langle {\bar \chi} \chi \rangle (m,\mu)
    = \frac{1}{V}\frac{\partial \ln Z(m,\mu)}{\partial m}\nn 
    \simeq
    \frac{1}{V}\frac{\ln Z(m+\Delta m,\mu)-\ln Z(m-\Delta m,\mu)}{2\Delta m},
    \label{eq:<chibarchi>}
\een
with $\Delta m=0.001$. 
The critical endpoint $m_c^{\chi}(N_\tau=8)$ is determined at the point where the magnitude of discontinuity $\langle {\bar \chi} \chi \rangle(m,\mu)$ associated with the first-order phase transition vanishes. Figure~\ref{fig:gap_nt8_chi} shows the $\mu$ dependence of $\langle {\bar \chi} \chi \rangle(m,\mu)$ with $\Delta \mu=0.001$ for several choices of $m$. 
Each curve exhibits a single clear jump at a certain value of $\mu$ for $m\in[1.90,2.025]$.  
The magnitude of discontinuity $\Delta \langle {\bar \chi} \chi \rangle(m)$ is evaluated by $\vert \langle {\bar \chi} \chi \rangle (m,\mu\rightarrow (\mu_++\mu_-)/2-0)-\langle {\bar \chi} \chi \rangle (m,\mu\rightarrow (\mu_++\mu_-)/2+0)\vert$, where $\mu_+$ and $\mu_-$ are chosen out of different phases, making $\vert \mu_+-\mu_-\vert$ as small as possible. 
The chiral condensate $\langle {\bar \chi} \chi \rangle (m,\mu\rightarrow (\mu_++\mu_-)/2\pm 0)$ is obtained by fitting seven data points closest to the transition point with a third-order polynomial in each phase. 
Again, we define $\mu_c^{\chi}=(\mu_++\mu_-)/2$, which is regarded as an estimate of the transition point.
The values of $\Delta \langle {\bar \chi} \chi \rangle(m)$ are summarized in Table~\ref{tab:gap_nt8}, where we find that $\mu_c^\chi$ agrees with $\mu_c^n$ for all $m$. 

\begin{figure}[htbp]
	\centering
	\includegraphics[width=0.85\hsize]{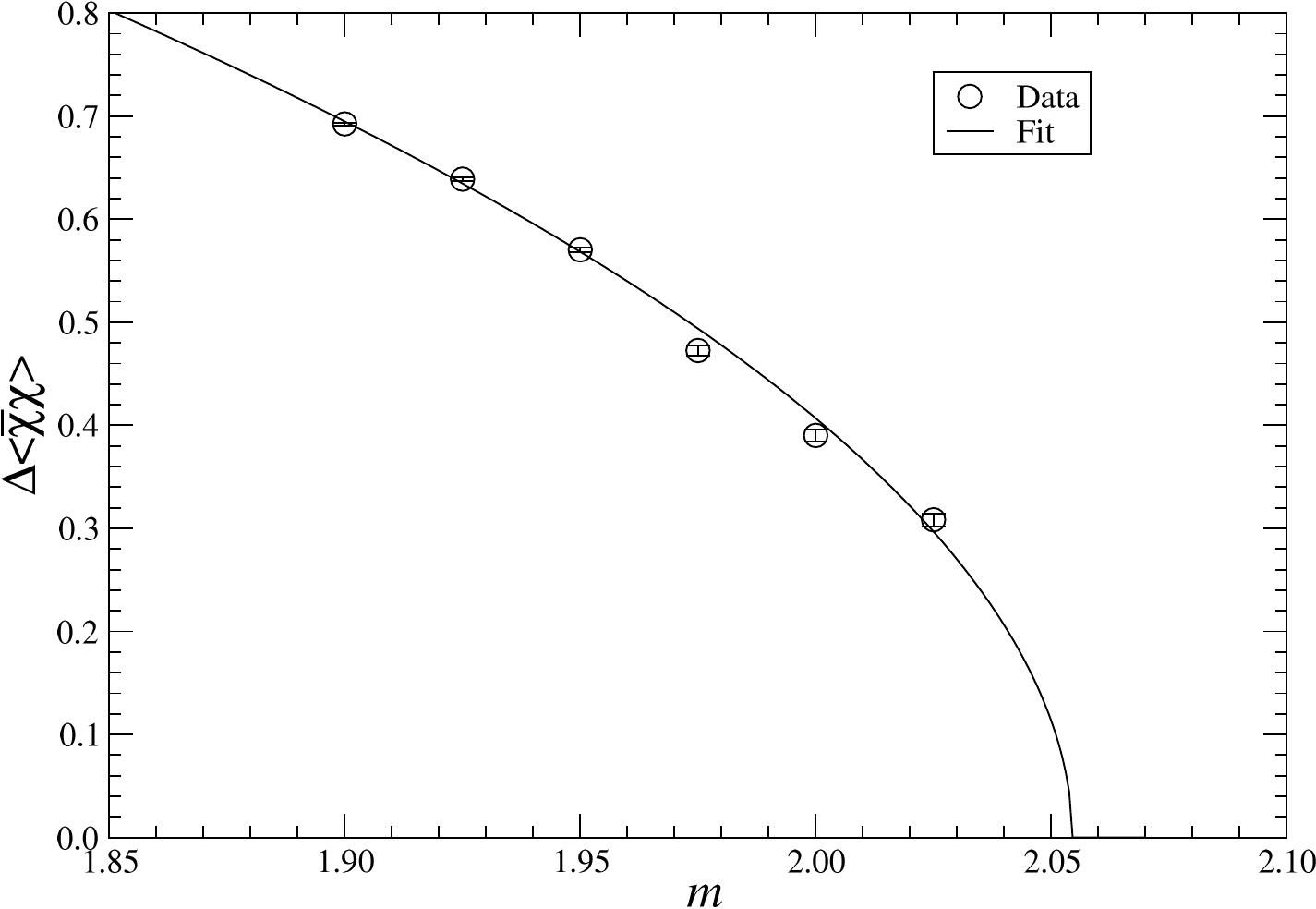}
	\caption{Fit of $\Delta \langle {\bar \chi} \chi \rangle(m)$ as a function of $m$. Solid curve denotes the fit result.}
  	\label{fig:fit_chi}
\end{figure}

The critical endpoint $m_c^\chi(N_\tau=8)$ is determined by fitting the data of $\Delta \langle {\bar \chi} \chi \rangle(m)$, assuming the functional form,
\begin{align}
    \Delta \langle {\bar \chi} \chi \rangle(m)=A_{\chi}\left(m_c^{\chi}(N_\tau=8)-m\right)^{p_{\chi}},
\end{align}
with $A_{\chi}$, $\mu_c$ and $p_{\chi}$ the free parameters.  Figure~\ref{fig:fit_chi} shows the fit result with 
\begin{align*}
    A_{\chi}=1.81(5),~~~m_c^{\chi}(N_\tau=8)=2.0545(34),~~~p_{\chi}=0.514(20).
\end{align*}
The value of $m_c^{\chi}(N_\tau=8)$ is consistent with $m_c^n(N_\tau=8)=2.075(23)$.
We also confirm that $\langle {\bar \chi} \chi \rangle$ at $m=2.10$ exhibits a smooth dependence on $\mu$ as shown in Fig.~\ref{fig:gap_nt8_chi}.

\subsection{Phase transition at zero temperature}
\label{subsec:nt=1024}

According to the MF analysis~\cite{Nishida:2003fb}, the phase transition at finite quark mass $m$ is expected to be of first order.
We investigate the phase transition at $m=2.07$ on a $1024^4$ lattice by calculating $\mu$ dependence of
the chiral condensate and the quark number density using Eqs.~(\ref{eq:<chibarchi>}) and (\ref{eq:<n>}) with $\Delta m=0.001$ and $\Delta \mu=0.001$. 
The thermodynamic potential is evaluated using the four-dimensional Grassmann ATRG algorithm with bond dimension $D=55$.

\begin{figure}[htbp]
	\centering
	\includegraphics[width=0.85\hsize]{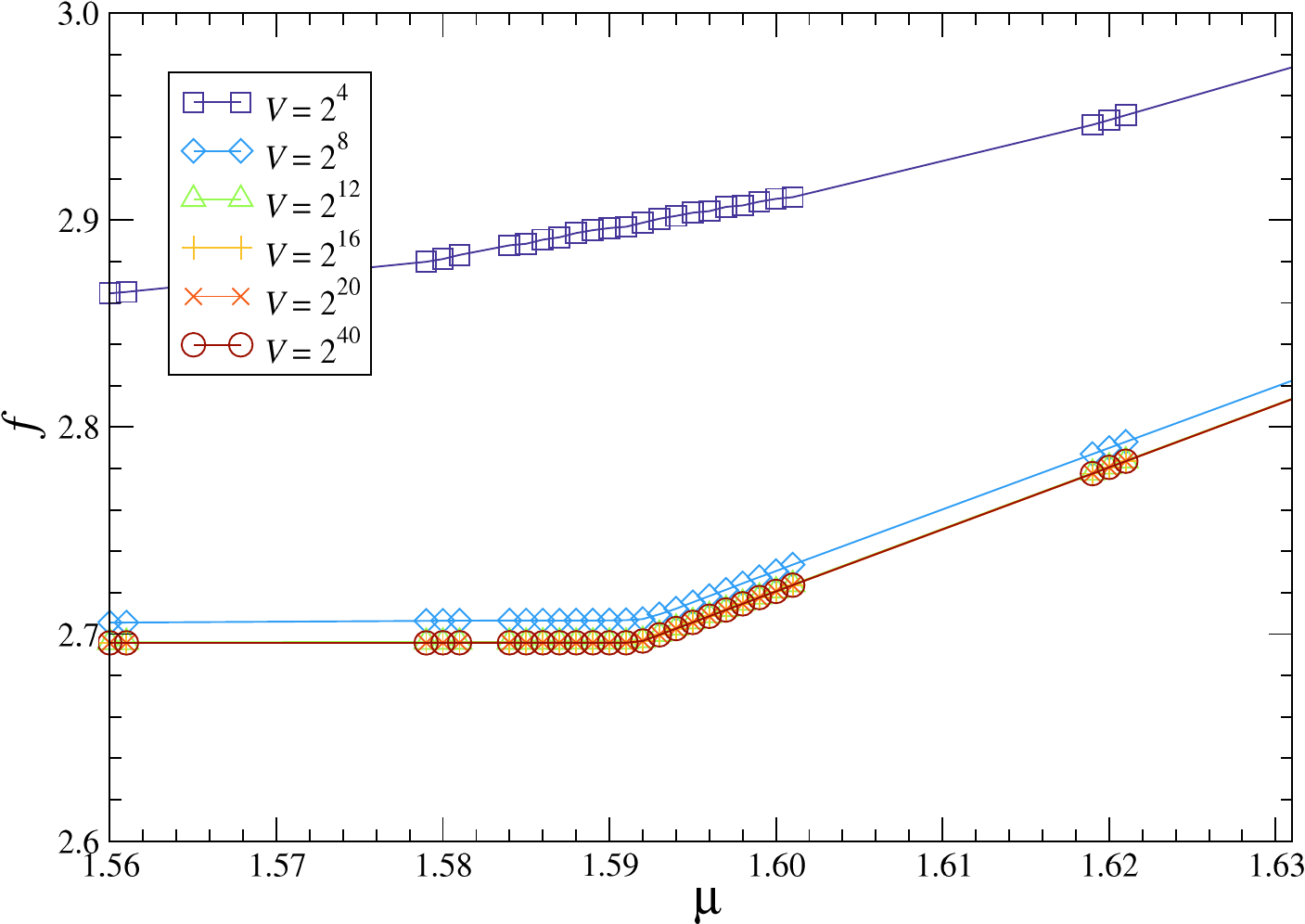}
	\caption{$\mu$ dependence of $f(m,\mu)$ at $m=2.07$. }
  	\label{fig:f_nt1024}
\end{figure}

\begin{figure}[htbp]
	\centering
	\includegraphics[width=0.8\hsize]{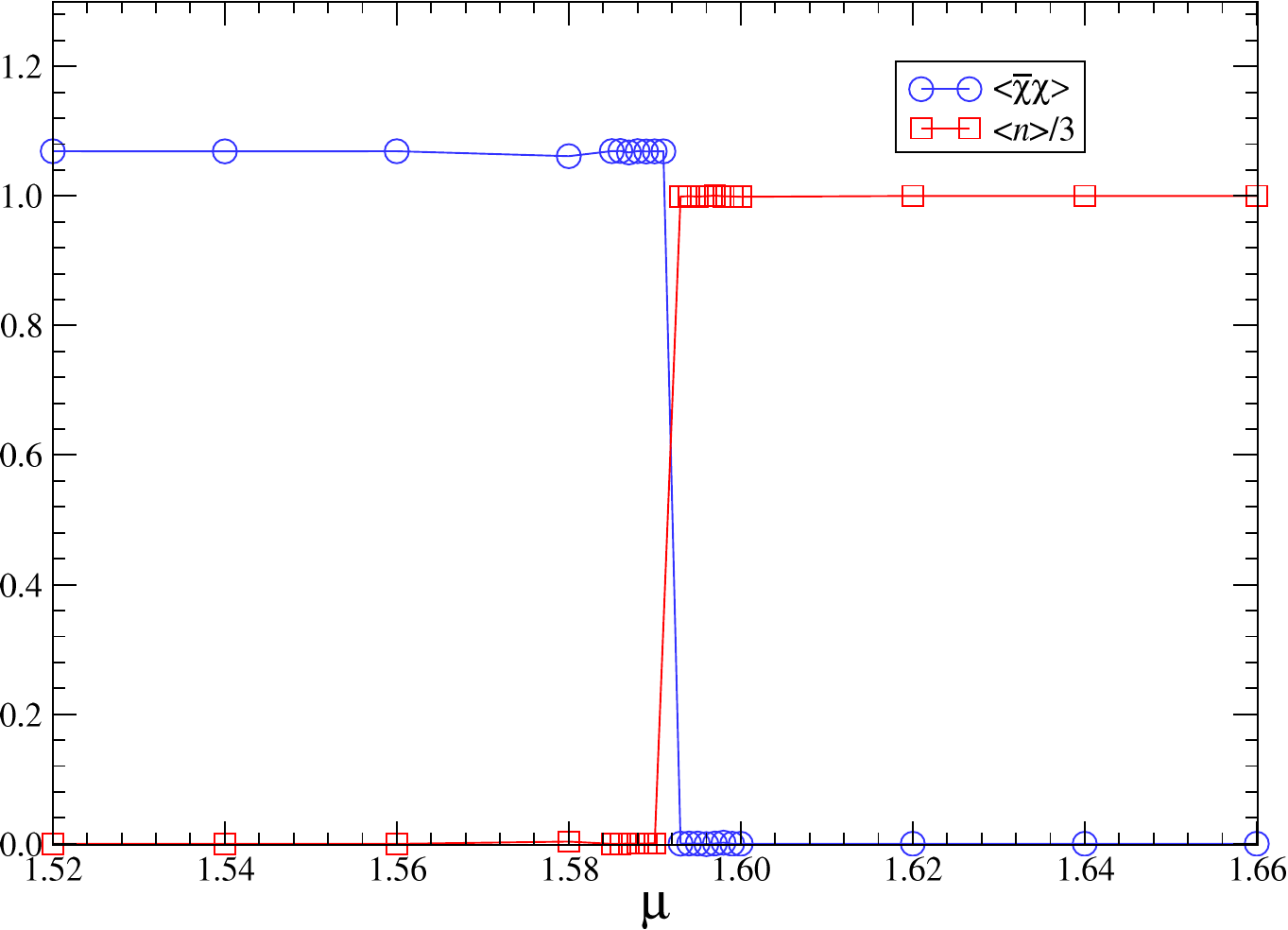}
	\caption{$\mu$ dependence of $\langle {\bar \chi} \chi\rangle$ and  $\langle n\rangle/3$ at $m=2.07$.}
  	\label{fig:mu-dep}
\end{figure}

Figure~\ref{fig:f_nt1024} shows the $\mu$ dependence of the thermodynamic potential.
We observe that a kink develops between $\mu=1.591$ and $1.592$ as the volume is increased. 
Compared to the $N_\tau=8$ case shown in the inset of Fig.~\ref{fig:f_nt8}, the thermodynamic limit is reached much more rapidly. 
Figure~\ref{fig:mu-dep} plots $\langle {\bar \chi} \chi\rangle$ and $\langle n\rangle$ at $m=2.07$ as a function of $\mu$. 
Both quantities exhibit a clear jump around $\mu\approx 1.591$, indicating a first-order phase transition. 
We observe that the transition is sharper than in the $N_\tau=8$ case shown in Figs.~\ref{fig:gap_nt8_n} and \ref{fig:gap_nt8_chi}. 
The transition point determined from $\langle \bar{\chi}\chi \rangle$ is $1.592(1)$, which is consistent with the value $1.5915(5)$ obtained from $\langle n \rangle$.
Note that $\langle n\rangle>3$ is not allowed due to the Pauli exclusion principle.

\section{Summary and outlook} 
\label{sec:summary}

We have investigated the phase structure of the (3+1)$d$ cold and dense QCD in the strong coupling limit. The critical endpoint at $N_\tau=8$ is determined by fitting the gap of chiral condensate and number density, yielding $m_c^\chi(N_\tau=8)=2.0545(34)$ and $m_c^n(N_\tau=8)=2.075(23)$, which agree with each other within the fitting error. 
The resulting critical quark mass lies between the values obtained from Monte Carlo simulation $m_c^n(N_\tau=8)=1.7(1)$ in the dual representation on finite lattices and that predicted by the MF analysis $m_c^n(N_\tau=8)\approx2.4$.
We also observe the first-order phase transition at $m=2.07$ on a $1024^4$ lattice, which is large enough to be regarded as the thermodynamic limit at zero temperature. This work ensures the ability of the TRG method to explore the cold and dense region of QCD, where the  Monte Carlo method is not allowed due to the complex action problem.

As a next step, it may be worthwhile to incorporate gauge fields at finite couplings. 
Recently, several attempts have been made to address how to incorporate dynamical non-Abelian gauge fields~\cite{Bazavov:2019qih,Fukuma:2021cni,Kuwahara:2022ubg,Asaduzzaman:2023pyz,Pai:2024tip,Pai:2025eia,Samberger:2026vpy}.
Combining these ideas with our approach will be an interesting future study.
We emphasize that there is no fundamental theoretical obstacle to embedding dynamical gauge fields into our Grassmann tensor network formulation, although the resulting computational complexity needs to be controlled.
It should be noted that there is no reason for the chiral and nuclear transitions to take place at the same chemical potential. It is interesting to investigate whether the introduction of finite coupling splits degenerate transitions or not.
Another possible direction is an extension to the Wilson quark. Although we use the Kogut--Susskind quark action to investigate the chiral transition in this work, the flavor symmetry of the Wilson quark would be useful in searching for the color-flavor locked phase in the high-density region.   

\begin{acknowledgments}
  Numerical calculation for the present work was carried out using the computational resources of SQUID provided by Osaka University through the HPCI System Research Project (Project ID: hp250120). We also used Yukawa-21 at Yukawa Institute Computer Facility in Kyoto University and the supercomputer Pegasus under the Multidisciplinary Cooperative Research Program of Center for Computational Sciences, University of Tsukuba. 
  Y.S. acknowledges support from the Graduate Program on Physics for the Universe (GP-PU), Tohoku University, and from JSPS KAKENHI (Grant-in-Aid for JSPS Fellows) Grant Number 25KJ0537.
  SA acknowledges the support from JSPS KAKENHI Grant Numbers JP23K13096 and JP25H01510, and the Top Runners in Strategy of Transborder Advanced Researches (TRiSTAR) program conducted as the Strategic Professional Development Program for Young Researchers by the MEXT. 
  This work is supported in part by Grants-in-Aid for Scientific Research from the Ministry of Education, Culture, Sports, Science and Technology (MEXT) (Nos. 24H00214, 24H00940).
\end{acknowledgments}
\clearpage
\appendix
\section{Explicit form of the fundamental Grassmann tensor}
\label{app:Gtn}
Here we give an explicit expression for the local tensor component $\mathcal{T}$. 
We first rewrite Eq.~\eqref{eq:local_tensor} in a form suitable for integrating out the link variables $U_\nu$,
\begin{align}\label{eq:local_tensor2}
    \mathcal{T}=&\sum_{\{i,j\}}\int \left(\prod_\nu dU_\nu\right)  d\chi^1 d\bar{\chi}^1 d\chi^2 d\bar{\chi}^2 d\chi^3 d\bar{\chi}^3 e^{-m\bar{\chi}^1\chi^1}e^{-m\bar{\chi}^2\chi^2}e^{-m\bar{\chi}^3\chi^3}\nn\\
&\quad\times\left[\prod_\nu\left(-\frac{\eta_\nu(n)}{2}e^{\mu\delta_{\nu,4}} \right)^{i^1_\nu+i^{2}_\nu+i^{3}_\nu}\left(-\frac{\eta_\nu(n)}{2}e^{-\mu\delta_{\nu,4}} \right)^{j^1_\nu+j^{2}_\nu+j^{3}_\nu}\right]\nn\\
&\quad\times\left[\prod_\nu \left(\sum_{a_\nu}\bar{\zeta}^3_\nu U^{3a_\nu}_\nu \chi^{a_\nu}\right)^{i'^3_\nu}\right]\left[\prod_\nu \left(\sum_{b_\nu}\bar{\zeta}^2_\nu U^{2b_\nu}_\nu \chi^{b_\nu}\right)^{i'^2_\nu}\right]\left[\prod_\nu \left(\sum_{c_\nu}\bar{\zeta}^1_\nu U^{1c_\nu}_\nu \chi^{c_\nu}\right)^{i'^1_\nu}\right]\nn\\
&\quad\times\left[\prod_\nu\left(\xi^1_\nu\chi^1\right)^{j^1_\nu}\right]\left[\prod_\nu\left(\xi^2_\nu\chi^2\right)^{j^2_\nu}\right]\left[\prod_\nu\left(\xi^3_\nu\chi^3\right)^{j^3_\nu}\right]\nn\\
&\quad\times\left[\prod_\nu \left(\sum_{d_\nu}\bar{\chi}^{d_\nu} U^{d_\nu3\dagger}_\nu\bar{\xi}^3_\nu  \right)^{j'^3_\nu}\right]\left[\prod_\nu  \left(\sum_{e_\nu}\bar{\chi}^{e_\nu} U^{e_\nu2\dagger}_\nu\bar{\xi}^2_\nu  \right)^{j'^2_\nu}\right]\left[\prod_\nu\left(\sum_{f_\nu}\bar{\chi}^{f_\nu} U^{f_\nu1\dagger}_\nu\bar{\xi}^1_\nu  \right)^{j'^1_\nu}\right]\nn\\
&\quad\times\left[\prod_\nu\left(\bar{\chi}^1\zeta^1_\nu\right)^{i^1_\nu}\right]\left[\prod_\nu\left(\bar{\chi}^2\zeta^2_\nu\right)^{i^2_\nu}\right]\left[\prod_\nu\left(\bar{\chi}^3\zeta^3_\nu\right)^{i^3_\nu}\right],
\end{align}
where $i'^{1}_{\nu}, i'^{2}_{\nu}, j'^{2}_{\nu}, j'^{1}_{\nu}, 
 j^{1}_{\nu}, j^{2}_{\nu}, i^{1}_{\nu}, i^{2}_{\nu} \in \{0,1\},(\nu=1,2,3,4)$ are the exponents arising from the Taylor expansion of the Grassmann variables in Eq.~\eqref{eq:local_tensor}, and auxiliary
integer variables $a_{\nu}, b_{\nu}, c_{\nu}, d_{\nu}, e_{\nu}, f_{\nu}\in \{1,2,3\},(\nu=1,2,3,4)$ which correspond to the color indices of the link variables. The index summation $\sum_{\{i,j\}}$ denotes $\prod_\nu\sum_{i^1_\nu,i^2_\nu,i^3_\nu,j^1_\nu,j^2_\nu,j^3_\nu,i'^1_\nu,i'^2_\nu,i'^3_\nu,j'^1_\nu,j'^2_\nu,j'^3_\nu}$.
To separate the link variables from the Grassmann variables, we employ the trick introduced in Ref.~\cite{Sugimoto:2025vui}. 
We define $g(i)=\tfrac{1}{3}\delta_{i,0}+\delta_{i,1}$, so that
\begin{align}
    \left(\sum_{a_\nu}\bar{\zeta}^3_\nu U^{3a_\nu}_\nu \chi^{a_\nu}\right)^{i'^3_\nu}
    &= g(i'^3_\nu) \sum_{a_\nu=1}^3 
       \left[(\bar{\zeta}^3_\nu)^{i'^3_\nu}\,
             (U^{3a_\nu}_\nu)^{i'^3_\nu}\,
             (\chi^{a_\nu})^{i'^3_\nu}\right],
\end{align}
and the same relation for terms including $b_{\nu}, c_{\nu}, d_{\nu}, e_{\nu}, f_{\nu}$.
By this transformation, we can analytically integrate out $U_\nu$ as follows:
\begin{align}
    &F^{a_\nu b_\nu c_\nu d_\nu e_\nu f_\nu}_{i'^3_\nu i'^2_\nu i'^1_\nu j'^3_\nu j'^2_\nu j'^1_\nu}\nn\\
    &=\int dU_\nu \Bigl(U_\nu^{3a_\nu}\Bigr)^{i'^3_\nu}\Bigl(U_\nu^{2b_\nu}\Bigr)^{i'^2_\nu}\Bigl(U_\nu^{1c_\nu}\Bigr)^{i'^1_\nu}\Bigl(U_\nu^{d_\nu3\dagger}\Bigr)^{j'^3_\nu}\Bigl(U_\nu^{e_\nu 2\dagger}\Bigr)^{j'^2_\nu}\Bigl(U_\nu^{f_\nu 1\dagger}\Bigr)^{j'^1_\nu}\nn\\
    &\qquad\times g(i'^3_\nu)g(i'^2_\nu)g(i'^1_\nu)g(j'^3_\nu)g(j'^2_\nu)g(j'^1_\nu)\nn\\
    &=\Big[
\delta_{0,i'^3_\nu}\delta_{0,i'^2_\nu}\delta_{0,i'^1_\nu}\delta_{0,j'^3_\nu}\delta_{0,j'^2_\nu}\delta_{0,j'^1_\nu}\nn\\
    &\quad+ \frac{1}{3}\delta_{1,i'^3_\nu}\delta_{0,i'^2_\nu}\delta_{0,i'^1_\nu} \delta_{1,j'^3_\nu}\delta_{0,j'^2_\nu}\delta_{0,j'^1_\nu}\delta^{a_\nu,d_\nu}\nn\\
&\quad+ \frac{1}{3}\delta_{0,i'^3_\nu}\delta_{1,i'^2_\nu}\delta_{0,i'^1_\nu} \delta_{0,j'^3_\nu}\delta_{1,j'^2_\nu}\delta_{0,j'^1_\nu}\delta^{b_\nu,e_\nu}\nn\\
&\quad+ \frac{1}{3}\delta_{0,i'^3_\nu}\delta_{0,i'^2_\nu}\delta_{1,i'^1_\nu} \delta_{0,j'^3_\nu}\delta_{0,j'^2_\nu}\delta_{1,j'^1_\nu}\delta^{c_\nu,f_\nu}\nn\\
&\quad
+\delta_{1,i'^3_\nu}\delta_{1,i'^2_\nu}\delta_{1,j'^3_\nu}\delta_{1,j'^2_\nu}\delta_{0,i'^1_\nu}\delta_{0,j'^1_\nu}
\left(\frac{1}{8}\delta^{a_\nu,d_\nu}\delta^{b_\nu,e_\nu}-\frac{1}{24}\delta^{a_\nu,e_\nu}\delta^{b_\nu,d_\nu}\right)\nn\\
&\quad
+\delta_{1,i'^3_\nu}\delta_{1,i'^1_\nu}\delta_{1,j'^3_\nu}\delta_{1,j'^1_\nu}\delta_{0,i'^2_\nu}\delta_{0,j'^2_\nu}
\left(\frac{1}{8}\delta^{a_\nu,d_\nu}\delta^{c_\nu,f_\nu}-\frac{1}{24}\delta^{a_\nu,f_\nu}\delta^{c_\nu,d_\nu}\right)\nn\\
&\quad
+\delta_{1,i'^2_\nu}\delta_{1,i'^1_\nu}\delta_{1,j'^2_\nu}\delta_{1,j'^1_\nu}\delta_{0,i'^3_\nu}\delta_{0,j'^3_\nu}
\left(\frac{1}{8}\delta^{b_\nu,e_\nu}\delta^{c_\nu,f_\nu}-\frac{1}{24}\delta^{b_\nu,f_\nu}\delta^{c_\nu,e_\nu}\right)\nn\\
&\quad
+\delta_{1,i'^3_\nu}\delta_{1,i'^2_\nu}\delta_{1,i'^1_\nu}\delta_{1,j'^3_\nu}\delta_{1,j'^2_\nu}\delta_{1,j'^1_\nu}
\Big[\frac{7}{120}\delta^{a_\nu,d_\nu}\delta^{b_\nu,e_\nu}\delta^{c_\nu,f_\nu}\nn\\
&\qquad\qquad
-\frac{1}{40}\big(\delta^{a_\nu,e_\nu}\delta^{b_\nu,d_\nu}\delta^{c_\nu,f_\nu}
+\delta^{a_\nu,d_\nu}\delta^{b_\nu,f_\nu}\delta^{c_\nu,e_\nu}
+\delta^{a_\nu,f_\nu}\delta^{b_\nu,e_\nu}\delta^{c_\nu,d_\nu}\big)\nn\\
&\qquad\qquad
+\frac{1}{60}\big(\delta^{a_\nu,e_\nu}\delta^{b_\nu,f_\nu}\delta^{c_\nu,d_\nu}
+\delta^{a_\nu,f_\nu}\delta^{b_\nu,d_\nu}\delta^{c_\nu,e_\nu}\big)\Big]\nn\\
&\quad
+\delta_{1,i'^3_\nu}\delta_{1,i'^2_\nu}\delta_{1,i'^1_\nu}\delta_{0,j'^3_\nu}\delta_{0,j'^2_\nu}\delta_{0,j'^1_\nu}\;\frac{1}{3!}\eps^{321}\eps^{a_\nu b_\nu c_\nu}\nn\\
&\quad
+\delta_{0,i'^3_\nu}\delta_{0,i'^2_\nu}\delta_{0,i'^1_\nu}\delta_{1,j'^3_\nu}\delta_{1,j'^2_\nu}\delta_{1,j'^1_\nu}\;\frac{1}{3!}\eps^{d_\nu e_\nu f_\nu}\eps^{321}
\Big]g(i'^3_\nu)g(i'^2_\nu)g(i'^1_\nu)g(j'^3_\nu)g(j'^2_\nu)g(j'^1_\nu),
\end{align}
where $\eps$ denotes the Levi-Civita symbol. 
The tensor $\mathcal{T}$ is now expressed as
\begin{align}
    \mathcal{T}&=\sum_{\{i,j\}}\int d\chi^1 d\bar{\chi}^1 d\chi^2 d\bar{\chi}^2 d\chi^3 d\bar{\chi}^3 e^{-m\bar{\chi}^1\chi^1}e^{-m\bar{\chi}^2\chi^2}e^{-m\bar{\chi}^3\chi^3}\left(\prod_\nu \sum_{a_\nu,b_\nu,c_\nu,d_\nu,e_\nu,f_\nu}\right)\left(\prod_\nu F^{a_\nu b_\nu c_\nu d_\nu e_\nu f_\nu}_{i'^3_\nu i'^2_\nu i'^1_\nu j'^3_\nu j'^2_\nu j'^1_\nu}\right)\nn\\
    &\times \left[\prod_\nu\left(\bar{\zeta}^3_\nu\chi^{a_\nu}\right)^{i'^3_\nu}\right]\left[\prod_\nu\left(\bar{\zeta}^2_\nu\chi^{b_\nu}\right)^{i'^2_\nu}\right]\left[\prod_\nu\left(\bar{\zeta}^1_\nu\chi^{c_\nu}\right)^{i'^1_\nu}\right]\left[\prod_\nu\left(\xi^1_\nu\chi^1\right)^{j^1_\nu}\right]\left[\prod_\nu\left(\xi^2_\nu\chi^2\right)^{j^2_\nu}\right]\left[\prod_\nu\left(\xi^3_\nu\chi^3\right)^{j^3_\nu}\right]\nn\\
    &\times  \left[\prod_\nu \left(\bar{\chi}^{d_\nu}\bar{\xi}^3_\nu\right)^{j'^3_\nu}\right]\left[\prod_\nu \left(\bar{\chi}^{e_\nu}\bar{\xi}^2_\nu\right)^{j'^2_\nu}\right]\left[\prod_\nu\left(\bar{\chi}^{f_\nu}\bar{\xi}^1_\nu\right)^{j'^1_\nu}\right]\left[\prod_\nu\left(\bar{\chi}^1\zeta^1_\nu\right)^{i^1_\nu}\right]\left[\prod_\nu\left(\bar{\chi}^2\zeta^2_\nu\right)^{i^2_\nu}\right]\left[\prod_\nu\left(\bar{\chi}^3\zeta^3_\nu\right)^{i^3_\nu}\right].\label{eq:T_after_integral_U}
\end{align}
To integrate $\chi,\bar{\chi}$, we should take care of the ordering of the Grassmann variables. In Eq.~\eqref{eq:T_after_integral_U}, the order of the $\chi^1,\chi^2,\chi^3$ and $\bar{\chi}^1,\bar{\chi}^2,\bar{\chi}^3$ depends on $i'^1_\nu,i'^2_\nu,i'^3_\nu,j'^1_\nu,j'^2_\nu,j'^3_\nu,a_\nu,b_\nu,c_\nu,d_\nu,e_\nu,f_\nu$. As in the case of the $SU(2)$~\cite{Sugimoto:2025vui}, redefining $F^{a_\nu b_\nu c_\nu d_\nu e_\nu f_\nu}_{i'^3_\nu i'^2_\nu i'^1_\nu j'^3_\nu j'^2_\nu j'^1_\nu}$ as 
\begin{align}
    &\tilde{F}^{a_\nu b_\nu c_\nu d_\nu e_\nu f_\nu}_{i'^3_\nu i'^2_\nu i'^1_\nu j'^3_\nu j'^2_\nu j'^1_\nu}\nn\\
    &=\Big[
\delta_{0,i'^3_\nu}\delta_{0,i'^2_\nu}\delta_{0,i'^1_\nu}\delta_{0,j'^3_\nu}\delta_{0,j'^2_\nu}\delta_{0,j'^1_\nu}\nn\\
    &\quad+ \frac{1}{3}\delta_{1,i'^3_\nu}\delta_{0,i'^2_\nu}\delta_{0,i'^1_\nu} \delta_{1,j'^3_\nu}\delta_{0,j'^2_\nu}\delta_{0,j'^1_\nu}\delta^{a_\nu,d_\nu}\nn\\
&\quad+ \frac{1}{3}\delta_{0,i'^3_\nu}\delta_{1,i'^2_\nu}\delta_{0,i'^1_\nu} \delta_{0,j'^3_\nu}\delta_{1,j'^2_\nu}\delta_{0,j'^1_\nu}\delta^{b_\nu,e_\nu}\nn\\
&\quad+ \frac{1}{3}\delta_{0,i'^3_\nu}\delta_{0,i'^2_\nu}\delta_{1,i'^1_\nu} \delta_{0,j'^3_\nu}\delta_{0,j'^2_\nu}\delta_{1,j'^1_\nu}\delta^{c_\nu,f_\nu}\nn\\
&\quad
+\delta_{1,i'^3_\nu}\delta_{1,i'^2_\nu}\delta_{1,j'^3_\nu}\delta_{1,j'^2_\nu}\delta_{0,i'^1_\nu}\delta_{0,j'^1_\nu}
\left(\frac{1}{8}\delta^{a_\nu,d_\nu}\delta^{b_\nu,e_\nu}+\frac{1}{24}\delta^{a_\nu,e_\nu}\delta^{b_\nu,d_\nu}\right)\nn\\
&\quad
+\delta_{1,i'^3_\nu}\delta_{1,i'^1_\nu}\delta_{1,j'^3_\nu}\delta_{1,j'^1_\nu}\delta_{0,i'^2_\nu}\delta_{0,j'^2_\nu}
\left(\frac{1}{8}\delta^{a_\nu,d_\nu}\delta^{c_\nu,f_\nu}+\frac{1}{24}\delta^{a_\nu,f_\nu}\delta^{c_\nu,d_\nu}\right)\nn\\
&\quad
+\delta_{1,i'^2_\nu}\delta_{1,i'^1_\nu}\delta_{1,j'^2_\nu}\delta_{1,j'^1_\nu}\delta_{0,i'^3_\nu}\delta_{0,j'^3_\nu}
\left(\frac{1}{8}\delta^{b_\nu,e_\nu}\delta^{c_\nu,f_\nu}+\frac{1}{24}\delta^{b_\nu,f_\nu}\delta^{c_\nu,e_\nu}\right)\nn\\
&\quad
+\delta_{1,i'^3_\nu}\delta_{1,i'^2_\nu}\delta_{1,i'^1_\nu}\delta_{1,j'^3_\nu}\delta_{1,j'^2_\nu}\delta_{1,j'^1_\nu}
\Big[\frac{7}{120}\delta^{a_\nu,d_\nu}\delta^{b_\nu,e_\nu}\delta^{c_\nu,f_\nu}\nn\\
&\qquad\qquad
-\frac{1}{40}\big(\delta^{a_\nu,e_\nu}\delta^{b_\nu,d_\nu}\delta^{c_\nu,f_\nu}
+\delta^{a_\nu,d_\nu}\delta^{b_\nu,f_\nu}\delta^{c_\nu,e_\nu}
+\delta^{a_\nu,f_\nu}\delta^{b_\nu,e_\nu}\delta^{c_\nu,d_\nu}\big)\nn\\
&\qquad\qquad
+\frac{1}{60}\big(\delta^{a_\nu,e_\nu}\delta^{b_\nu,f_\nu}\delta^{c_\nu,d_\nu}
+\delta^{a_\nu,f_\nu}\delta^{b_\nu,d_\nu}\delta^{c_\nu,e_\nu}\big)\Big]\nn\\
&\quad
+\delta_{1,i'^3_\nu}\delta_{1,i'^2_\nu}\delta_{1,i'^1_\nu}\delta_{0,j'^3_\nu}\delta_{0,j'^2_\nu}\delta_{0,j'^1_\nu}\;\frac{1}{3!}\eps^{321}|\eps^{a_\nu b_\nu c_\nu}|\nn\\
&\quad
+\delta_{0,i'^3_\nu}\delta_{0,i'^2_\nu}\delta_{0,i'^1_\nu}\delta_{1,j'^3_\nu}\delta_{1,j'^2_\nu}\delta_{1,j'^1_\nu}\;\frac{1}{3!}|\eps^{d_\nu e_\nu f_\nu}|\eps^{321}
\Big]g(i'^3_\nu)g(i'^2_\nu)g(i'^1_\nu)g(j'^3_\nu)g(j'^2_\nu)g(j'^1_\nu),
\end{align}
we are free from the troublesome sign factors.
Finally, we obtain $\mathcal{T}$ by defining the coefficient tensor $T$ as in the following form,
\begin{align}
    \mathcal{T}&=\sum_{\{i,j\}}T_{I_1I_2I_3I_4I'_1I'_2I'_3I'_4}\nn\\
    &\qquad\times\left[(\zeta^1_1)^{i^1_1}(\zeta^2_1)^{i^2_1}(\zeta^3_1)^{i^3_1}(\xi^1_1)^{j^1_1}(\xi^2_1)^{j^2_1}(\xi^3_1)^{j^3_1}\right]
        \left[(\zeta^1_2)^{i^1_2}(\zeta^2_2)^{i^2_2}(\zeta^3_2)^{i^3_2}(\xi^1_2)^{j^1_2}(\xi^2_2)^{j^2_2}(\xi^3_2)^{j^3_2}\right]\nn\\
        &\qquad\times\left[(\zeta^1_3)^{i^1_3}(\zeta^2_3)^{i^2_3}(\zeta^3_3)^{i^3_3}(\xi^1_3)^{j^1_3}(\xi^2_3)^{j^2_3}(\xi^3_3)^{j^3_3}\right]
        \left[(\zeta^1_4)^{i^1_4}(\zeta^2_4)^{i^2_4}(\zeta^3_4)^{i^3_4}(\xi^1_4)^{j^1_4}(\xi^2_4)^{j^2_4}(\xi^3_4)^{j^3_4}\right]\nn\\
    &\qquad\times\left[(\bar{\xi}^3_4)^{j'^3_4}(\bar{\xi}^2_4)^{j'^2_4}(\bar{\xi}^1_4)^{j'^1_4}(\bar{\zeta}^3_4)^{i'^3_4}(\bar{\zeta}^2_4)^{i'^2_4}(\bar{\zeta}^1_4)^{i'^1_4}\right]\left[(\bar{\xi}^3_3)^{j'^3_3}(\bar{\xi}^2_3)^{j'^2_3}(\bar{\xi}^1_3)^{j'^1_3}(\bar{\zeta}^3_3)^{i'^3_3}(\bar{\zeta}^2_3)^{i'^2_3}(\bar{\zeta}^1_3)^{i'^1_3}\right]\nn\\
    &\qquad\times\left[(\bar{\xi}^3_2)^{j'^3_2}(\bar{\xi}^2_2)^{j'^2_2}(\bar{\xi}^1_2)^{j'^1_2}(\bar{\zeta}^3_2)^{i'^3_2}(\bar{\zeta}^2_2)^{i'^2_2}(\bar{\zeta}^1_2)^{i'^1_2}\right]\left[(\bar{\xi}^3_1)^{j'^3_1}(\bar{\xi}^2_1)^{j'^2_1}(\bar{\xi}^1_1)^{j'^1_1}(\bar{\zeta}^3_1)^{i'^3_1}(\bar{\zeta}^2_1)^{i'^2_1}(\bar{\zeta}^1_1)^{i'^1_1}\right].\label{eq:T_after_integral_chi}
\end{align}
Here we have introduced shorthand notation $I_\nu=(i^1_\nu,i^2_\nu,i^3_\nu,j^1_\nu,j^2_\nu,j^3_\nu)$
and $I'_\nu=(i'^1_\nu,i'^2_\nu,i'^3_\nu,j'^1_\nu,j'^2_\nu,j'^3_\nu)$.
For simplicity, we define the auxiliary variable,
\begin{align}
\theta_\nu(p)&=\delta^{p,a_\nu}i'^3_\nu+\delta^{p,b_\nu}i'^2_\nu+\delta^{p,c_\nu}i'^1_\nu+j^p_\nu,\\
\eta_\nu(q)&=\delta^{q,d_\nu}j'^3_\nu+\delta^{q,e_\nu}j'^2_\nu+\delta^{q,f_\nu}j'^1_\nu+i^q_\nu.
\end{align}
The explicit form of the coefficient tensor $T$, given by integrating $\chi,\bar{\chi}$, is
\begin{align}
        &T_{I_1I_2I_3I_4I'_1I'_2I'_3I'_4}\nonumber\\
    &=\sum_{\{i,j\}}\left[\prod_\nu\left(-\frac{\eta_\nu(n)}{2}e^{\mu\delta_{\nu,4}} \right)^{i^1_\nu+i^{2}_\nu+i^{3}_\nu}\left(-\frac{\eta_\nu(n)}{2}e^{-\mu\delta_{\nu,4}} \right)^{j^1_\nu+j^{2}_\nu+j^3_{\nu}}\right]\left(\prod_\nu \sum_{a_\nu,b_\nu,c_\nu,d_\nu,e_\nu,f_\nu}\right)\left(\prod_\nu \tilde{F}^{a_\nu b_\nu c_\nu d_\nu e_\nu f_\nu}_{i'^3_\nu i'^2_\nu i'^1_\nu j'^3_\nu j'^2_\nu j'^1_\nu}\right)\nonumber\\
    &\quad\times \Big(\delta_{1, \sum_\nu \theta_\nu(1)}\delta_{1, \sum_\nu \theta_\nu(2)}\delta_{1, \sum_\nu \theta_\nu(3)}\delta_{1, \sum_\nu \eta_\nu(1)}\delta_{1, \sum_\nu \eta_\nu(2)}\delta_{1, \sum_\nu \eta_\nu(3)}\nonumber\\
    &\qquad\quad+m\delta_{0, \sum_\nu \theta_\nu(1)}\delta_{1, \sum_\nu \theta_\nu(2)}\delta_{1, \sum_\nu \theta_\nu(3)}\delta_{0, \sum_\nu \eta_\nu(1)}\delta_{1, \sum_\nu \eta_\nu(2)}\delta_{1, \sum_\nu \eta_\nu(3)}\nonumber\\
    &\qquad\quad+m\delta_{1, \sum_\nu \theta_\nu(1)}\delta_{0, \sum_\nu \theta_\nu(2)}\delta_{1, \sum_\nu \theta_\nu(3)}\delta_{1, \sum_\nu \eta_\nu(1)}\delta_{0, \sum_\nu \eta_\nu(2)}\delta_{1, \sum_\nu \eta_\nu(3)}\nonumber\\
    &\qquad\quad+m\delta_{1, \sum_\nu \theta_\nu(1)}\delta_{1, \sum_\nu \theta_\nu(2)}\delta_{0, \sum_\nu \theta_\nu(3)}\delta_{1, \sum_\nu \eta_\nu(1)}\delta_{1, \sum_\nu \eta_\nu(2)}\delta_{0, \sum_\nu \eta_\nu(3)}\nonumber\\
    &\qquad\quad-m^2 \delta_{1, \sum_\nu \theta_\nu(1)}\delta_{0, \sum_\nu \theta_\nu(2)}\delta_{0, \sum_\nu \theta_\nu(3)}\delta_{1, \sum_\nu \eta_\nu(1)}\delta_{0, \sum_\nu \eta_\nu(2)}\delta_{0, \sum_\nu \eta_\nu(3)}\nonumber\\
    &\qquad\quad-m^2 \delta_{0, \sum_\nu \theta_\nu(1)}\delta_{1, \sum_\nu \theta_\nu(2)}\delta_{0, \sum_\nu \theta_\nu(3)}\delta_{0, \sum_\nu \eta_\nu(1)}\delta_{1, \sum_\nu \eta_\nu(2)}\delta_{0, \sum_\nu \eta_\nu(3)}\nonumber\\
    &\qquad\quad-m^2 \delta_{0, \sum_\nu \theta_\nu(1)}\delta_{0, \sum_\nu \theta_\nu(2)}\delta_{1, \sum_\nu \theta_\nu(3)}\delta_{0, \sum_\nu \eta_\nu(1)}\delta_{0, \sum_\nu \eta_\nu(2)}\delta_{1, \sum_\nu \eta_\nu(3)}\nonumber\\
    &\qquad\quad-m^3\delta_{0, \sum_\nu \theta_\nu(1)}\delta_{0, \sum_\nu \theta_\nu(2)}\delta_{0, \sum_\nu \theta_\nu(3)}\delta_{0, \sum_\nu \eta_\nu(1)}\delta_{0, \sum_\nu \eta_\nu(2)}\delta_{0, \sum_\nu \eta_\nu(3)}\Big)\nonumber\\
    &\qquad\times(-1)^{R_{I_1I_2I_3I_4I'_1I'_2I'_3I'_4}}.\label{eq:coef_T}
\end{align}
The tensor $R_{I_1I_2I_3I_4I'_1I'_2I'_3I'_4}$ accounts for the fermionic sign factors arising from reordering the Grassmann variables from the ordering in Eq.~\eqref{eq:T_after_integral_U} to that in Eq.~\eqref{eq:T_after_integral_chi}.

Since each subscript $I_\nu$ and $I'_\nu$ of the initial coefficient tensor $T$ contains six Grassmann indices, the size of the coefficient tensor scales as $64^8$.
Therefore, it is impractical to construct it explicitly by nested for-loops due to the memory and computational costs.
To this end, we exploit the sparsity of $T$.
The Grassmann integration over $\chi$ and $\bar{\chi}$ imposes strong constraints on the nonzero elements of $T$.
In particular, $i'_\nu$ and $j_\nu$ correspond to the occupation numbers of $\chi^1,\chi^2,\chi^3$, while $i_\nu$ and $j'_\nu$ correspond to those of $\bar{\chi}^1,\bar{\chi}^2,\bar{\chi}^3$.
Since each of $\chi^1,\chi^2,\chi^3$ (and similarly $\bar{\chi}^1,\bar{\chi}^2,\bar{\chi}^3$) can appear at most once, a necessary condition for $T$ to be nonzero is\footnote{One can impose stricter constraints. However, the present framework is sufficient for this model.}
\begin{align}
\sum_{\nu=1}^{4}\sum_{c=1}^{3}\left(i'^{c}_{\nu}+j^{c}_{\nu}\right)\leq 3,\qquad\sum_{\nu=1}^{4}\sum_{c=1}^{3}\left(i^{c}_{\nu}+j'^{c}_{\nu}\right)\leq 3.
\label{eq:constraints}
\end{align}
Accordingly, we construct $T$ in a sparse format and evaluate only the index combinations allowed by Eq.~\eqref{eq:constraints}. In the initial truncated singular value decomposition step for constructing the five-leg tensor in ATRG, we use the sparse eigensolver \texttt{ARPACK}~\cite{doi:10.1137/1.9780898719628}.
\bibliographystyle{JHEP}
\bibliography{bib/formulation,bib/algorithm,bib/discrete,bib/grassmann,bib/continuous,bib/gauge,bib/real_time,bib/review,bib/for_this_paper}

\end{document}